\documentclass[preprint,amsmath,amssymb,aps,showkeys,showpacs]{revtex4}
\usepackage[colorlinks=true, pdfstartview=FitV, linkcolor=blue, citecolor=red, urlcolor=magenta]{hyperref}
\usepackage{graphicx}
\usepackage{latexsym}
\usepackage{amsmath}
\usepackage{amsfonts}
\usepackage{amssymb}
\usepackage{verbatim}
\usepackage{dcolumn}
\usepackage{amssymb}
\usepackage{bm}
\usepackage{float}
\usepackage{subfigure}
\usepackage[english]{babel}
\newcommand{\be}{\begin{equation}}
\newcommand{\ee}{\end{equation}}
\newcommand{\bea}{\begin{eqnarray}}
\newcommand{\eea}{\end{eqnarray}}

\graphicspath{{Figures/}}

\begin{document}

\newcommand{\NITK}{
\affiliation{Department of Physics, National Institute of Technology Karnataka, Surathkal  575025, India}
}
\newcommand{\GK}{
\affiliation{Department of Physics, Government College, Kasaragod,
Kerala, 671123, India}
}
\newcommand{\TKM}{\affiliation{
Department of Physics, T.K.M College of Arts and Science, Kollam, Kerala 691005, India.
}}

\title{Ruppeiner geometry, $P-V$ criticality and interacting microstructures of black holes in dRGT massive gravity}

\author{T. K. Safir}
\email{stkphy@gmail.com}
\NITK
\TKM
\author{C. L. Ahmed Rizwan}
\email{ahmedrizwancl@gmail.com}
\GK
\author{Deepak Vaid}
\email{dvaid79@gmail.com}
\NITK

    \begin{abstract}
    
    We probe the microstructure of the dRGT massive black hole in an anti-de Sitter background. The calculations are performed in an extended phase space with pressure and volume taken as fluctuation variables. We analyze the microstructure by exploiting the Ruppeiner geometry, where the thermodynamic curvature scalar is constructed via the adiabatic compressibility. The nature of the curvature scalar along the coexistence line of small (SBH) and large (LBH) black holes is investigated. In the microscopic interaction, we observe that the SBH phase behaves as an anyonic gas and the LBH phase is analogous to a boson gas. Further, we study the effect of graviton mass on the underlying microstructure of the black hole.
    
    \end{abstract}

\keywords{Massive gravity, Extended black hole thermodynamics, Black hole microstructures}

\maketitle

\section{Introduction}
The quest for the quantum nature of gravity exploiting the properties of black holes has been active for a long time. In the early 1970s, Hawking proved that the area of the event horizon would never decrease under any physical processes \citep{PhysRevLett.26.1344}. Considering this property as a characteristic of thermodynamic entropy, Bekenstein argued that black holes could be attributed to entropy, which is related to the event horizon area \citep{Bekenstein1973}. This was the first hint which indicated that black holes behave like thermodynamic objects. Soon, four laws were established for a general black hole in stationary, asymptotically flat spacetime \citep{Bardeen:1973gs}. The tools of thermodynamics, which are used extensively in ordinary systems, are now applied to black hole systems. Accordingly, further studies in black hole thermodynamics showed that thermally stable black holes exist only in anti-de Sitter (AdS) spacetime. A milestone in this direction was provided by Hawking and Page in the 1980s. They found a phase transition in Schwarzchild- AdS black hole known as Hawking-Page transition, which is between radiation and large black hole phases \citep{Hawking1983}. Later, the inconsistency between the first law and the Smarr relation led to the identification of the cosmological constant as the pressure term in the thermodynamics studies of black hole\citep{Kastor:2009wy}. Subsequently, it was observed that the $P-V$ criticality of charged AdS black holes shows resemblance with that of the Van der Waals (vdW) fluid \citep{Kubiznak2012}. The thermodynamics of AdS black holes in the extended phase space with a $PdV$ term exhibits a first-order phase transition similar to the liquid/gas transition in vdW fluid. In AdS black holes, the transition is between a small black hole (SBH) and a large black hole (LBH) phases. Also, a variety of other thermodynamic properties of vdW fluids like the Joule-Thomson effect and heat engine were found in AdS black holes \citep{Okcu:2016tgt,Johnson:2014yja}.\\


From the statistical point of view, the thermodynamics of a system always demands a microscopic description. Even though the thermodynamics of black holes is widely studied, understanding the microscopic structure of the black hole was a challenging problem. In this regard, the Ruppeiner geometry approach is a handy tool that extracts certain aspects of microscopic information from macroscopic properties. In this approach, a metric is defined on the thermodynamic state-space, which measures the distance between nearby fluctuating states \citep{Ruppeiner:1995zz, Ruppeiner:2008kd}. The Ruppeiner curvature scalar (R), which can be calculated from the metric, provides crucial details about phase transitions and the nature of interactions in the microstructure of the system under consideration. The important information obtained from this construction are; (1) nature of the microstructure interaction, (2) strength of the interaction, and (3) critical behavior. The sign of curvature scalar specifies the type of interactions. For a non-interacting system like an ideal gas, the curvature scalar vanishes. In an interacting system, the positive and negative values of the scalar curvature represent repulsive and attractive interactions, respectively. Also, the magnitude of the Ruppeiner curvature scalar measures the strength of interaction. An added feature of the construction is that, for the system with critical behavior, the curvature scalar shows divergence at the critical point. The singularity of the curvature scalar is related to the singular nature of response functions near the critical point. The Ruppeiner geometric method was found effective in describing the microstructure details of a variety of known systems in conventional thermodynamics \cite{Ruppeiner:1995zz, Janyszek_1990, Oshima_1999x, Mirza2008, PhysRevE.88.032123}. This success eventually led to the application of the Ruppeiner geometry method to a black hole systems \cite{Ruppeiner:2008kd}.\\

 A variety of approaches have been presented within the framework of Ruppeiner geometry to understand the microstructure of black holes \citep{Wei2015, Wei2019a, Wei2019b, Guo2019, Miao2017, Zangeneh2017, Wei:2019ctz, NaveenaKumara:2020biu, Kumara:2020ucr, Rizwan:2020bhp, Kumara:2020mvo,  Kumara:2019xgt, Xu:2019nnp, Chabab2018, Deng2017, Miao2019a, Chen2019, Du:2019poh, Dehyadegari2017, Xu:2019gqm, Miao:2018fke, Xu:2020gzm, Xu:2020ftx, Ghosh:2019pwy, Ghosh:2020kba, Yerra:2020oph, Wu:2020fij}. Recently, it was proposed that the thermodynamic curvature scalar can be constructed by taking the pressure and volume as fluctuating variables, and the curvature scalar is normalized by adiabatic compressibility \citep{Dehyadegari:2020ebz}. This construction makes an observation that strong repulsive interactions dominate among the microstructures of small black holes where the thermodynamic curvature diverges to positive infinity. Further, this method has been successfully applied to the Gauss-Bonnet AdS black hole spacetime \citep{NaveenaKumara:2020hov}. In this paper, we focus on the microscopic interactions of massive dRGT black holes using the curvature scalar normalized via adiabatic compressibility.\\ 

 Even though recent observations from LIGO/VIRGO collaborations confirm the predictions of the general theory of relativity, there are enough reasons to look beyond Einstein's theory of gravity. The cosmological constant problem, the hierarchy problem, and the accelerated expansion demand a modification to the General theory of relativity. The general theory of relativity is an unique massless spin two field theory. Among the several modified vrsions massive gravity theories explain the universes accelerated expansion without introducing a cosmological constant or dark energy. Massive gravity theories have a long and remarkable history. The initial studies were carried out by Fierz and Pauli in 1939\citep{Fierz:1939ix}. The proposed theory was linear and ghost-free but did not reduce to general relativity in the massless limit. Non-linear modifications of Fierz and Pauli's theory lead to "Boulware-Deser" ghost instability\citep{PhysRevD.6.3368}. Later,  de Rham, Gabadadze, and Tolley (dRGT) came up with a special class of non-linear massive gravity theory, which is "Boulware-Deser" ghost free\citep{PhysRevLett.106.231101}. Further, the thermodynamics of the black holes in massive gravity were widely investigated \citep{Cai:2014znn, Xu:2015rfa, Hendi:2017fxp, Hendi:2015bna, Mirza:2014xxa, Fernando:2016qhq, Ning:2016usb}.
    The van der Waals like feature of dRGT massive gravity black holes and other applications such as triple point, Reentrant phase transitions, heat engines, and throttling process were also studied \citep{Zou:2016sab, Liu:2018jld, Hendi:2017bys, Yerra:2020bfx, Lan:2019kak}. In addition, several works to probe the microstructure were also studied using various thermodynamic-geometry approaches \citep{Chabab:2019mlu, Wu:2020fij, Yerra:2020oph}.\\
    
 In this paper, we study the effect of graviton mass and different topologies on the microstructures of AdS black holes in massive dRGT theory. The article is organised in the following ways. We review the thermodynamics of black holes in massive gravity theory in section II. The following section discusses the Ruppeiner geometry and analyzes a few critical features. In section IV, we conclude the paper with certain remarks.
 
\section{Thermodynamics of Black Holes in massive Gravity}
In this section, we discuss the spacetime and thermodynamic structure of black holes in massive gravity theory. Here we consider dRGT non-linear massive gravity theory. In four-dimensional AdS space, the action for the Einstein-dRGT gravity coupled to a non-linear electromagnetic field reads as
\begin{equation}
S=\int d^4 x \sqrt{-g} \left[ \frac{1}{16 \pi}\left[ R+\frac{6}{l^2}+m^2\sum_{i=1}^4 c_i \ \mathcal{U}_i(g,f)\right ]-\frac{1}{4\pi } F_{\mu \nu}F^{\mu \nu}\right],
\end{equation} 
 where $F_{\mu \nu}=\partial_\mu A_\nu-\partial_\nu A_\mu $ is the electromagnetic field tensor with vector potential $A_\mu$, $l$ is AdS radius, $m$ is related to the graviton mass, and $c_i$ are coupling parameters. Further, $f_{\mu \nu}$ is a symmetric tensor as reference metric coupled to the space-time metric $g_{\mu \nu}$. Graviton interaction terms are represented by symmetric polynomials $\mathcal{U}_i$, and are obtained from a $4 \times 4 $ matrix $\mathcal{K}^\mu_\nu =\sqrt{g^{\mu \alpha} f_{\nu \alpha}}$, which have the following forms,
\begin{eqnarray*}
\begin{split}
\mathcal{U}_1 &= [\mathcal{K}]\\
\mathcal{U}_2 & = [\mathcal{K}]^2-[\mathcal{K}^2]\\
\mathcal{U}_3 & = [\mathcal{K}]^3-3[\mathcal{K}^2][\mathcal{K}]+2[\mathcal{K}^3]\\
\mathcal{U}_4 & = [\mathcal{K}]^4-6[\mathcal{K}^2][\mathcal{K}]^2+8[\mathcal{K}^3][\mathcal{K}]+3[\mathcal{K}^2]^2-6[\mathcal{K}^4]
\end{split}
\end{eqnarray*}
The solution to the above action for various horizon topologies are given by\citep{PhysRevD.95.021501,Cai:2014znn},

\begin{equation}
ds^2=-f(r)dt^2+\frac{1}{f(r)}dr^2+r^2h_{ij} dx_i dx_j,
\end{equation}
where $h_{ij}$ is the metric for two dimensional hypersurface.  The topological parameter ($k$) can take values $0,-1$ or $1$,  representing  planar, hyperbolic, and spherical topology, respectively. With the choice of reference metric $f_{\mu\nu}=diag(0,0,c_0^2 h_{ij})$, the values of $\mathcal{U}_i$ becomes $\mathcal{U}_1=\frac{2c_0}{r}$,~~ $\mathcal{U}_2=\frac{2c_0^2}{r^2},~~\mathcal{U}_3= \mathcal{U}_4= 0$. Now, the metric function reduces to,

\begin{equation}\label{metric_1}
f(r)=k-\frac{m_0}{r}-\frac{\Lambda r^2}{3} +\frac{q^2}{r^2}+m^2\left( \frac{c_0 c_1}{2}r +c_0^2 c_2\right),
\end{equation}
where integration constants $m_0$ and $q$ are related to  black hole mass and charge, respectively. $m$ is the parameter for graviton mass, and in the limiting case of  $m=0$ the spacetime reduces to Reissner- Nordstrom black hole solution.

With this quick review of spacetime, we now turn to its thermodynamics in the extended phase space. Here the cosmological constant $\Lambda$ is dynamic and is related to the pressure as $P=-
\Lambda /8\pi$, and its conjugate quantity gives the black hole volume (normalized) \citep{Kastor:2009wy}. In extended thermodynamics, the mass of the black hole is interpreted as enthalpy of the spacetime rather than energy. The Hawking temperature of the black hole is associated with the surface gravity $T=\frac{\kappa}{2\pi}$, with $\kappa=\frac{f'(r_h)}{2}$, where $r_h$ is the horizon radius. Another important thermodynamic parameter, the entropy of the black hole can be obtained from the Bekenstein area law. These thermodynamics quantities can easily be obtained as,

\begin{align*}
M=& \left( \frac{r_h}{2}(k +c_0^2 c_2 m^2)+\frac{c_0c_1 m^2 r_h}{2	}+\frac{8}{3}\pi P r_h^2 +\frac{q^2}{4r_h^2}\right),\\
T=& \left( 2P r_h+\frac{k+c_0^2c_2m^2}{4\pi r_h	}-\frac{q^2}{16\pi r_h^3} +\frac{c_0c_1 m^2}{4\pi}\right),\\ 
S=& \pi r_h^2.
\end{align*}
The first law of black hole mechanics can be readily written using the above quantities, using which we can calculate the remaining thermodynamic quantities. Thermodynamic volume is,
\begin{equation}
V=\left(\frac{\partial M}{\partial P}\right)_{S,Q}=\frac{4}{3}\pi r_h^3 =\frac{\pi}{6}v^3,
\end{equation} 
where $v=2r$ is the specific volume. The equation of state of the system is,
\begin{equation}
P=\frac{q^2}{2\pi v^4}-\frac{k+c_0^2c_2m^2}{2\pi v^2}+\frac{T}{v}-\frac{c_0 c_1 m^2}{4\pi v}. 
\end{equation}

The black hole exhibits a vdW like behaviour. The critical values  can be derived from the conditions,
\begin{equation}
\left( \frac{\partial P}{\partial v} \right)_T=0, \qquad \text{and} \qquad \left( \frac{\partial ^2 P}{\partial v^2} \right)_T=0,
\end{equation}
we obtain,
\begin{equation}
P_c=\frac{(k+m^2c_2c_0^2)^2}{24\pi q^2}, \qquad
v_c=\frac{\sqrt{6}\pi q}{(k+m^2c_2c_0^2)^{1/2}}, \qquad
T_c=\frac{2\left(k+m^2c_2c_0^2\right)^{(3/2)}}{3\sqrt{6}\pi q}+\frac{m^2 c_1c_0}{4\pi}.
\end{equation} 

\begin{figure}[t]
\centering
\includegraphics[scale=0.9]{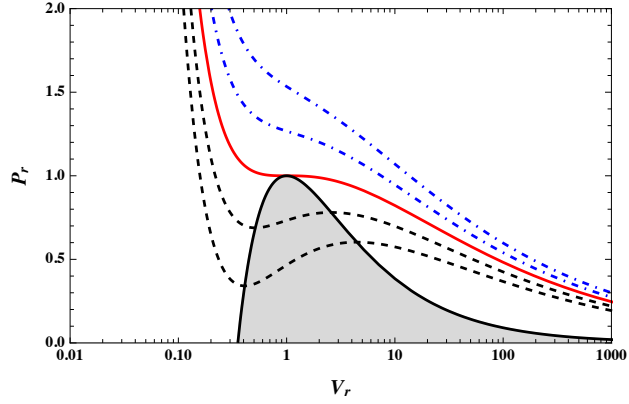}
\caption{$P-V$ isotherms of the massive dRGT-AdS black hole. The shaded region below the solid black line corresponds to the unstable states.  (Parameters are in reduced terms and the $x$ axis is in the log scale).   }
\label{fig1}
\end{figure}

Using the above critical values, we define the reduced thermodynamic quantities as,
\begin{equation}
P_r =\frac{P}{P_c}, \qquad T_r=\frac{T}{T_c}, \qquad v_r=\frac{v}{v_c}, \qquad V_r=\frac{V}{V_c}. 
\end{equation}
In the reduced parameter space, the equation of state reduces to,
\begin{align}\notag
P_r=\frac{8}{3v}\Bigg\{T_r\left[1+ \frac{c_0c1m^2 q}{16}\left(\frac{6}{k+c_0^2c_2m^2}\right)^{3/2}\right] \\ 
-\frac{c_0c_1 m^2q}{16}\left(\frac{6}{k+c_0^2c_2m^2}\right)^{3/2}\Bigg\}+\frac{1}{3v^4}-\frac{2}{v^2}.
\end{align}
In fact, we can write the  equation of state in terms of thermodynamic volume $V$ as, 
\begin{equation}
P= \frac{-4 {c_0}^2 {c_2} m^2-4 k}{8\ 6^{2/3} \left(\pi\right)^{1/3}  V^{2/3}}+\frac{16 \pi  T-4 {c_0} {c_1} m^2}{16 \left(6\right)^{1/3} \pi ^{2/3} \left(V\right)^{1/3}}+\frac{\left(\frac{\pi }{6}\right)^{1/3} q^2}{12 V^{4/3}}.
\end{equation}
The above equation $P(V,T)$ (since $r=\left( \frac{3V}{4\pi}\right)^{1/3}$ ) is often called the geometric form. Now, one can construct the Maxwell equal area law. The observed coexistence line determines the small-large black hole transition region. In the next sections, we examine the microstructure of the dRGT black hole using Ruppeiner geometry and present our observations in detail.

\section{Ruppeiner Geometry and Microstructure of Massive Black Hole}
In this section, we study the interacting microstructure of the black hole using Ruppeiner geometry method. Without compromising generality we will put $c_0=1$ and $m=1$ in Eq. (\ref{metric_1}). In the thermodynamic parameter space of pressure and entropy, the line element can be simplified to the following form,\citep{Dehyadegari:2020ebz} 
\begin{equation}
    dl^2=\frac{dS^2}{C_P}+\frac{V}{T}\kappa_s
dP^2,
\end{equation}
here the heat capacity $C_P = T( \frac{\partial S}{\partial T})_P$ and $\kappa_s=\frac{-1}{V}(\frac{\partial V}{\partial P})_S$. Considering the interdependence of entropy and volume in the case of a spherically symmetric black hole, the above line element can be written as \begin{equation}
    dl^2=\frac{1}{C_P}\left(\frac{\pi}{6V}\right)^{2/3} dV^2+\frac{V}{T}\kappa_S
dP^2
\end{equation}
Now, we have pressure and volume as the thermodynamic variables. Since the adiabatic compressibility $\kappa_S$ is a vanishing quantity similar to the heat capacity at a constant volume, we define a normalized thermodynamic curvature as $R_N=R \kappa_S$. 
By performing a direct calculation of the curvature scalar, we obtain the normalized curvature scalar $(R_N)$ of the dRGT black hole.
\begin{align}\notag
R_N = \frac{16 {V_r}^{2/3}}{y^{5/2} \left(P_r  {V_r}^{4/3}-2 {V_r}^{2/3}+1\right)^2 \left(36 {c_1} q {V_r}+\sqrt{6} y^{3/2} \left(3 P_r  {V_r}^{4/3}+6 {V_r}^{2/3}-1\right)\right)} \\ \notag \times \Bigg\{ 54 \sqrt{6} {c_1}^2 q^2 {V_r}^2+9 {c_1} q {V_r} y^{3/2} \left(9 P_r  {V_r}^{4/3}+6 {V_r}^{2/3}+1\right)\\ 
+\sqrt{6} \left(3 {V_r}^{2/3}-1\right) y^3 \left(9 P_r  {V_r}^{4/3}-6 {V_r}^{2/3}+5\right)\Bigg\}.
\end{align}
\begin{figure}[t]
\centering
\includegraphics[scale=0.9]{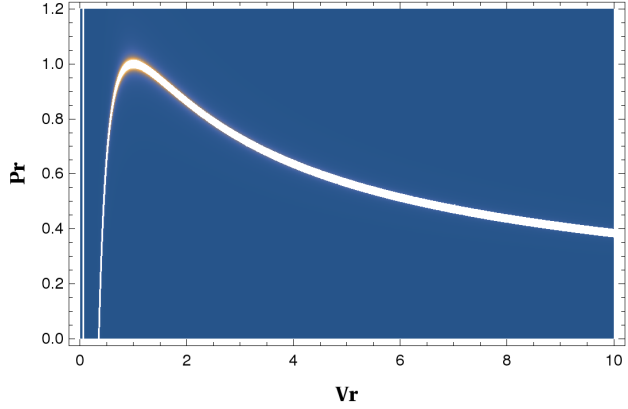}
\caption{The behaviour of  $R_N$ as a function of $P$ and $V$  }
\label{fig1}
\end{figure}

\begin{figure*}[t]
\centering
\subfigure[ref1][]{\includegraphics[scale=0.75]{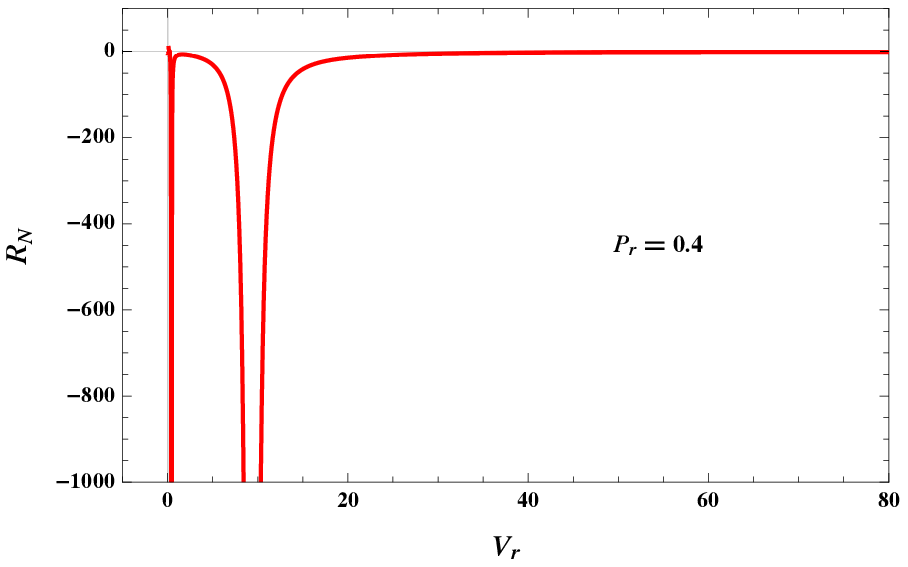}\label{RNVR1}}
\qquad
\subfigure[ref2][]{\includegraphics[scale=0.75]{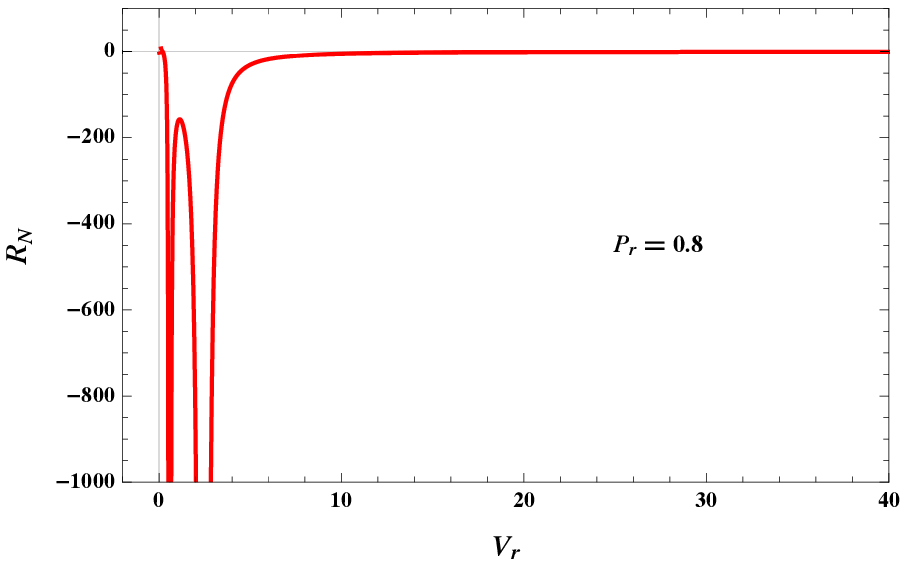}\label{RNVR2}}

\subfigure[ref1][]{\includegraphics[scale=0.75]{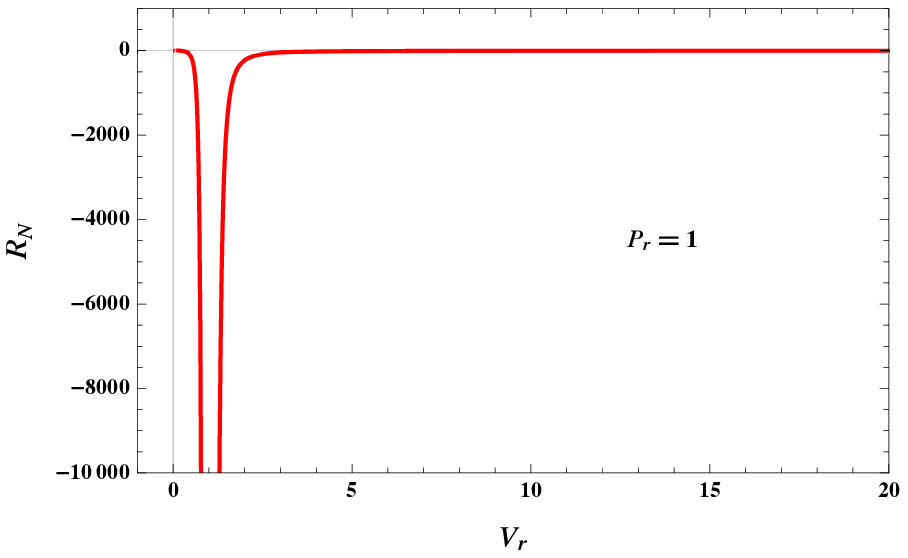}\label{RNVR3}}
\qquad
\subfigure[ref1][]{\includegraphics[scale=0.75]{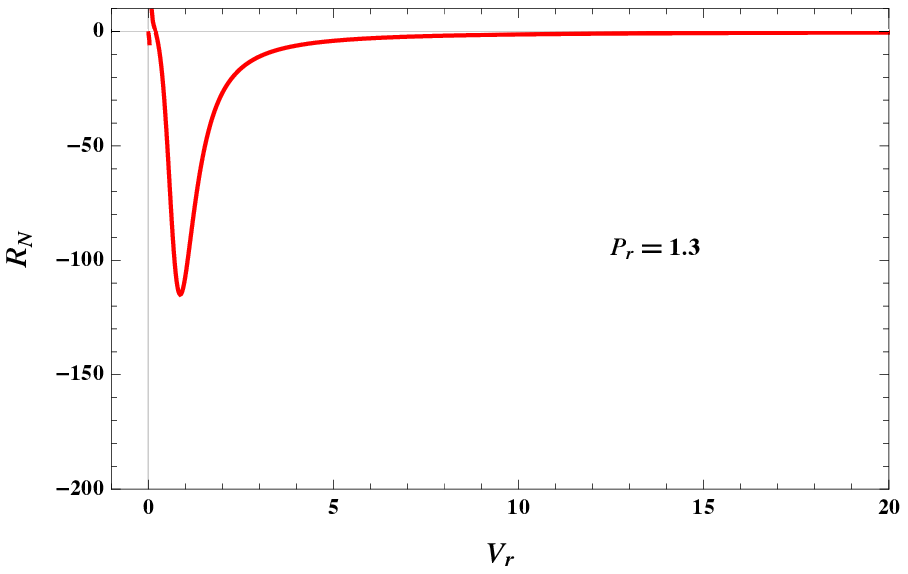}\label{RNVR4}}
\caption{The behaviour of the $R_N$ against the reduced volume $V_r$ at constant pressure. }
\label{RN}
\end{figure*}

\begin{figure*}[t]
\centering
\subfigure[][]{\includegraphics[scale=0.75]{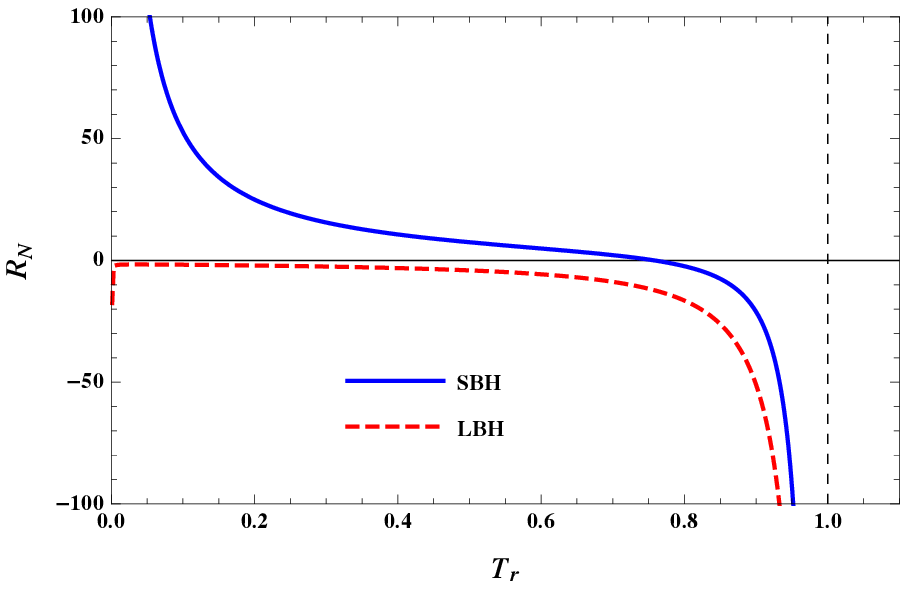}\label{MGRT}}
\qquad
\subfigure[][]{\includegraphics[scale=0.75]{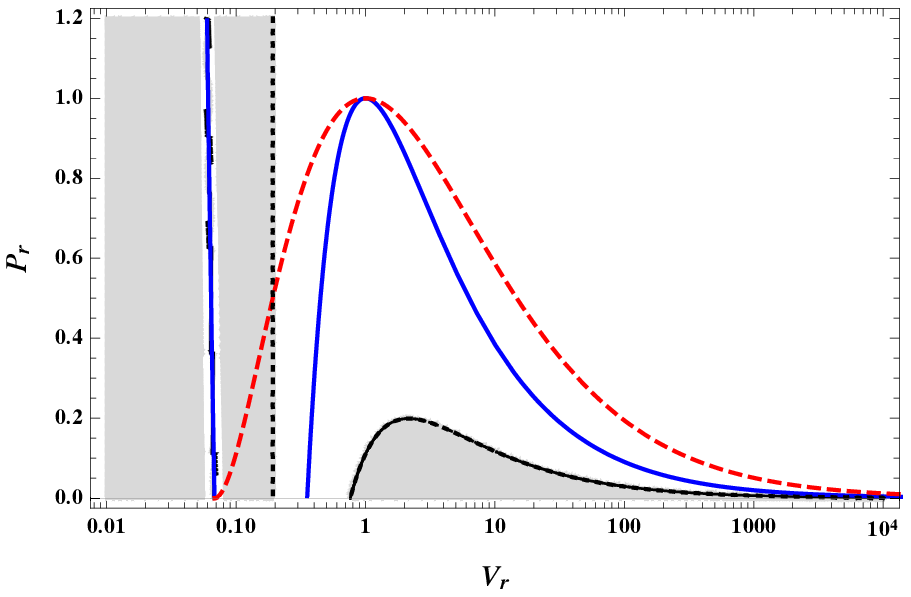}\label{RPV}}

\caption{Left: The behaviour of  $R_N$ along the coexistence curve. The red (dashed) and blue (solid) lines correspond to LBH and SBH, respectively. Right: The vanishing curve (black dot-dashed line) and diverging curve (blue solid line) of $R_N$ along with the coexistence curve (red dashed line). The shaded regions indicate positive $R_N$; otherwise, $R_N$ is negative (Parameters are in reduced terms, and the $x$ axis is in the log scale).}
\label{RT}
\end{figure*}

Here, the variable $y$ is defined as $y=k+c_2$. The behaviour of $R_N$ with reduced volume $V_r$ for a fixed parameter is studied in the figures \ref{RNVR1} to \ref{RNVR4}. For $P_r<1$, $R_N$ has two negative divergences. These divergences merge at $V_r=1$ for $P_r=1$. For $P_r>1$, these divergences of $R_N$ disappears.
The divergence of $R_N$ is along the curve is defined by,
\begin{equation}
    P_{div}=\frac{2V_r^{2/3}-1}{V_r^{4/3}} \qquad
    P_{div}=\frac{1-2V_r^{2/3}}{3V_r^{4/3}}-\frac{0.003061}{V_r^{1/3}}
\end{equation}
Here, the dominant interaction is attractive in nature because the $R_N$ is always negative.
In the figure \ref{RPV}, the red dashed curve represents the coexistence curve and the blue solid represents the divergent curve. The shaded region in the graph indicates the positive values of the normalized scalar curvature and the other remaining regions $R_N$ take negative values. SBH and LBH phases simultaneously coexist under the coexistence curve. From this figure, it is apparent that a certain range of volume in SBH has positive $R_N$, which implies the domination of repulsive interaction. The region where $R_N$ is negative signifies attractive microstructure interactions. 

 One can study the plot for $R_N$ along the coexistence curve for both SBH and LBH phases from critical temperature to zero. Here we observe that $R_N$ for both SBH and LBH diverges to $-\infty$ at the critical temperature. The figure shows that LBH possesses only negative values of $R_N$, and it gradually increases while approaching the critical temperature. However, for the SBH phase $R_N$ changes sign and become a positive value below a particular temperature. Also, $R_N$ goes to positive infinity as T tends to zero, where strong repulsive interaction dominates.
 
\begin{figure*}[h]
\centering
\subfigure[][]{\includegraphics[scale=0.75]{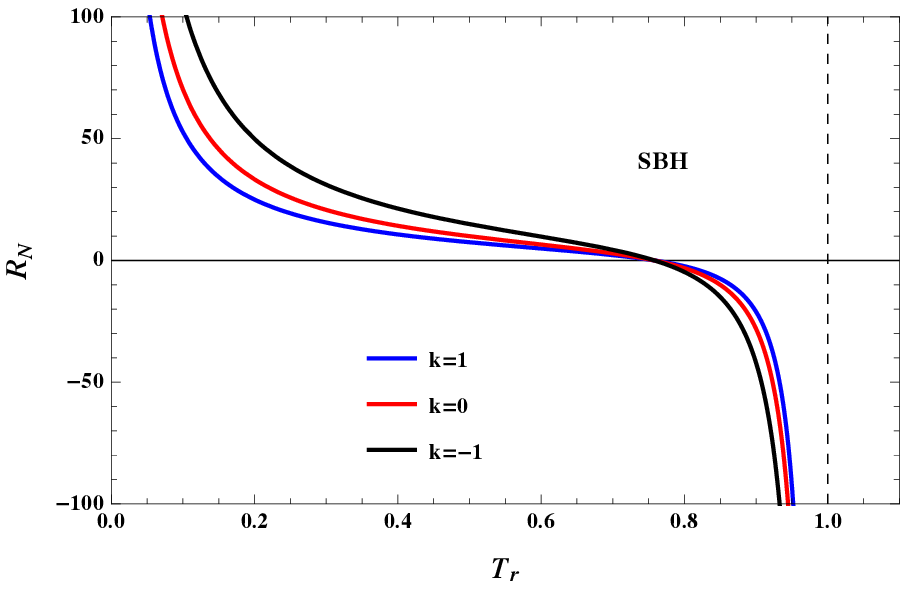}\label{RTksbh}}
\qquad
\subfigure[][]{\includegraphics[scale=0.75]{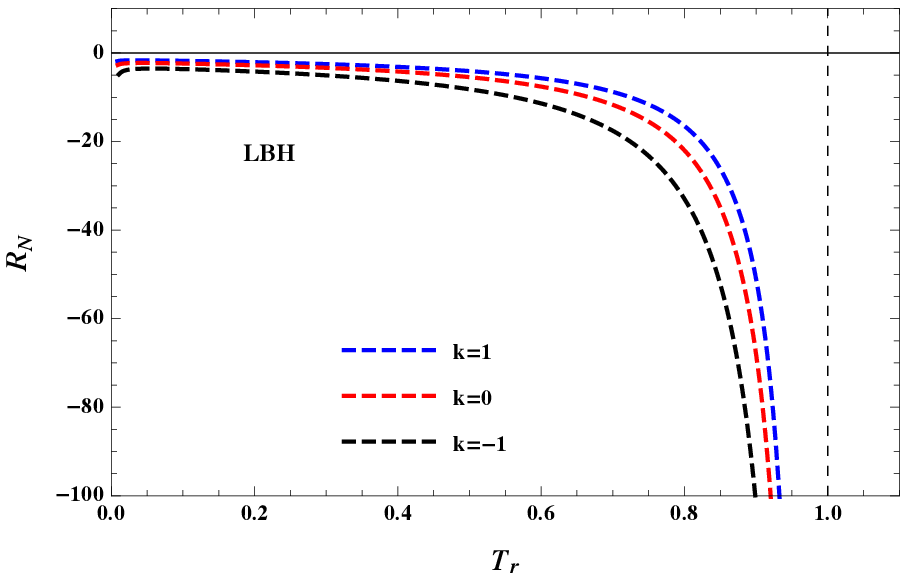}\label{RTklbh}}

\caption{The effect of parameter $k$ on the microstructure interactions. In the left SBH and in the right LBH are shown.}
\label{RTk}
\end{figure*}

 We also investigate the effect of graviton mass ($m=0,1,2$) and horizon topology on the micro-states of dRGT black holes for SBH and LBH phases. It is clear from the figure that the repulsive interaction of SBH becomes strongly repulsive as graviton mass increases. However, the attractive nature of the LBH phase becomes weaker as graviton mass increases. Figures \ref{RTksbh} and \ref{RTklbh} show the repulsive interaction SBH is stronger for spherical topology, followed by flat and hyperbolic topology. Hyperbolic topology has the most attractive LBH interactions, followed by flat and spherical topology.

\begin{figure*}[t]
\centering
\subfigure[][]{\includegraphics[scale=0.75]{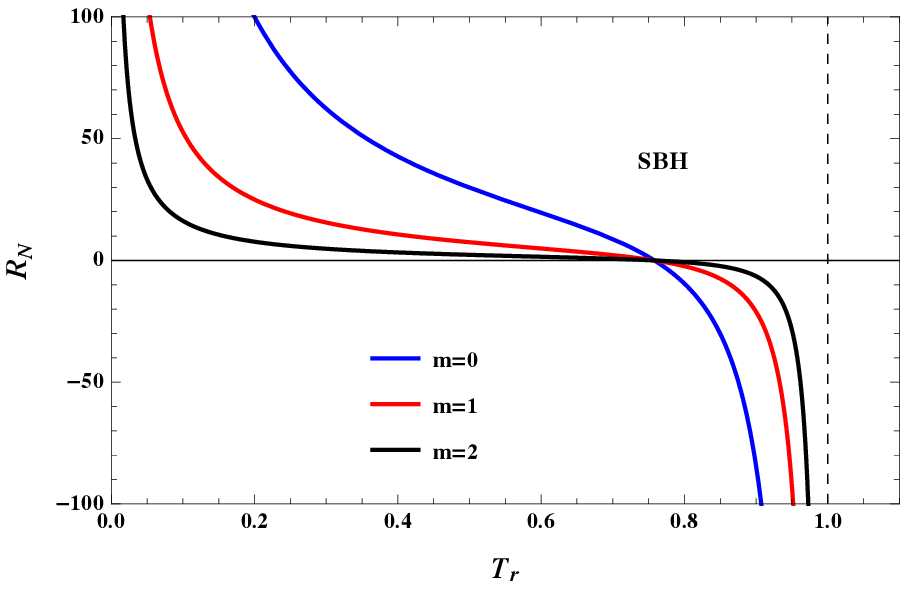}\label{RTmsbh}}
\qquad
\subfigure[][]{\includegraphics[scale=0.75]{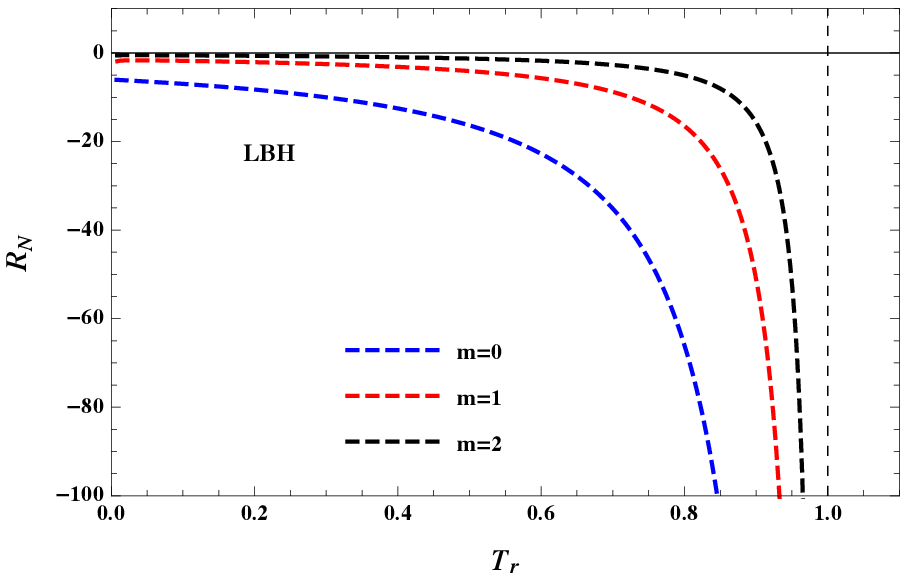}\label{RTmlbh}}
\caption{The effect of parameter $m$ on the microstructure interactions. In the left SBH and in the right LBH are shown. }
\label{RTm}
\end{figure*}

\section{Concluding Remarks}
\label{conclusion}
In this article, we have constructed the Ruppeiner geometry for an AdS black hole in dRGT massive gravity to probe its microstructure interactions. The underlying motivation for this study lies in the fact that the black hole phase transition is related to its microstructure details. The construction is carried out by defining a normalized curvature scalar $R_N$ in the parameter space of pressure $P$ and the volume $V$, which are the fluctuation coordinates, via the adiabatic compressibility $\kappa$. The phenomenological understanding of the nature and the strength of the microstructure follows from the observation of the behavior of $R_N$ along the coexistence line for the small-large black hole phase transition, which is a first-order transition with universal behavior. The validity of the construction of normalized curvature scalar is confirmed by looking at the divergence of the curvature scalar at the critical point of the phase transition. The study shows that the microstructure details of the dRGT massive black holes are similar to the charged AdS black holes; however, it is influenced by the parameters of the spacetime that govern the massive gravity background.\\

The small black hole phase has microstructures analogous to anyon gas with attractive and repulsive interactions. The result is interesting due to the presence of a repulsive interaction at some parameter space, which differs from the microstructure properties of a vdW fluid (where we have only dominant attractive interactions), though the phase transition properties are akin to both black hole and VdW systems. In other words, the universality in the critical behavior does not imply the similarity in the microstructure interactions. We note that the repulsive interaction is suppressed by the graviton mass $m$. The effect of the topology on the repulsive interaction was also investigated, which shows a trend of stronger to weaker in the order $k=-1,0,1$. The LBH exhibits only attractive interactions all over the parameter space, which is similar to a boson gas. In both the small and large black hole phases, the attractive interactions of the black hole microstructure have the same dependence on the spacetime parameters $m$ and $k$. The effect of charge on the microstructure is also explored, which we find negligible.

\acknowledgments
We thank Naveena Kumara A. and Fairoos C. for suggestions and discussions.

  \bibliography{BibTex}
  
\end{document}